\documentclass{emulateapj}
\usepackage{color}
\slugcomment{Submitted to Astrophysical Journal}

\shorttitle{Growing Disks in DM Halos}
\shortauthors{Berentzen and Shlosman}

\def\gtorder{\mathrel{\raise.3ex\hbox{$>$}\mkern-14mu
    \lower0.6ex\hbox{$\sim$}}}
\def\ltorder{\mathrel{\raise.3ex\hbox{$<$}\mkern-14mu
    \lower0.6ex\hbox{$\sim$}}}

\newcommand{\Bl}{ \left( }
\newcommand{\Br}{ \right)}

\begin{document}

\title{Growing Live Disks Within Cosmologically Assembling Asymmetric Halos:\\
                      Washing out the Halo Prolateness}

\author{Ingo Berentzen and Isaac Shlosman}

\affil{
Department of Physics and Astronomy,
University of Kentucky,
Lexington, KY 40506-0055,
USA \\ 
email: {\tt iberent@pa.uky.edu, shlosman@pa.uky.edu}
}

\begin{abstract}
We study the growth of galactic disks in live triaxial dark matter (DM) halos. The halos 
have been assembled using the constrained realizations method and evolved from the 
linear regime using cosmological simulations. The `seed' disks have been inserted at 
redshift $z=3$ and increased in mass tenfold over various time periods, $\sim 1-3$~Gyr, with
the halo responding quasi-adiabatically to this process. We follow the 
dynamical and secular evolution of the disk-halo system and analyze changes in the 
most important parameters, like three-dimensional DM shapes, stellar (disk) and DM (halo)
radial density profiles, stellar bar development, etc. We find that a growing disk is 
responsible for washing out the halo prolateness and for diluting its flatness over a period 
of time comparable to the disk growth. Moreover, we find that a disk which contributes more to 
the overall rotation curve in the system is also more efficient in axisymmetrizing the 
halo, without accelerating the halo figure rotation. The observational corollary is that 
the maximal disks probably reside in nearly axisymmetric halos, while disks whose 
rotation is dominated by the halo at all radii are expected to reside in more prolate halos. 
The halo shape is sensitive to the final disk mass, but is independent of how the seed 
disk is introduced into the system --- abruptly or quasi-adiabatically. It is weakly 
sensitive to the timescale of the disk growth. We also expect that the massive disks are 
subject to a bar instability, while light disks have this instability damped by the halo 
triaxiality. Implications of these results to the cosmological evolution of disks 
embedded in asymmetric halos are discussed and so are the corollaries for the observed
fraction of stellar bars. Finally, the halo responds to the stellar bar by developing a 
gravitational wake --- a `ghost' bar of its own which is almost in-phase with that 
in the disk.
\end{abstract}


\section{Introduction}

A great deal of effort has been invested during the last couple of decades 
in understanding the large-scale structure formation in the Universe. While
a substantial progress has been made in following the hierarchical buildup
of dark matter (DM) halos, the structure development on subgalactic scales,
such as the growth of disks, bulges, central black holes, etc., is only now
being investigated. In this paper we study the buildup of stellar disks 
immersed in evolving DM halos of an arbitrary shape obtained in cosmological 
simulations and the concurrent symbiotic evolution of the disk-halo system. 
Specifically, we analyze the change in the halo triaxiality to the growing 
disk and its feedback onto the disk evolution.

Beyond the active and passive adjustments of the halo to the growing disk 
and {\it vice 
versa}, the disk is prone to a number of dynamical instabilities, such as the
bar instability, which can alter the evolution of the system in the most
profound way. A strong and massive stellar bar will facilitate the angular
momentum and mass redistribution within the disk and between the disk and
the halo (e.g., Athanassoula 2003). The efficiency of this process is not yet
fully understood but is known to depend on a number of parameters
such as the dispersion velocities in the halo, its mass concentration, etc.

First, from the most general dynamical considerations, the coupled dynamical 
evolution 
of a disk-halo system is expected to depend on the prevailing shape of the DM
halo and its embedded disk as well as on the ability of the disk to develop
a strong stellar bar. While DM halos form on the average as highly triaxial 
(i.e., prolate and flat) systems in cosmological numerical simulations (e.g.,
Bullock 2002 for a review), they appear nearly axisymmetric (i.e., oblate) 
in the local Universe (e.g., Rix \& Zaritsky 1995; Merrifield 2002). What 
process(es) drive this axisymmetrization of the halo?

Second, the halo radial density profile can have an effect
on its coupling to the growing disk, especially on the efficiency of the angular
momentum transfer to the halo. It was argued that cuspy halos (i.e., Navarro,
Frenk \& White 1997, hereafter NFW) provide a substantial drag on the
stellar bars (e.g., Weinberg 1985; but see Sellwood 2006) and the feedback is 
capable of leveling off the density cusp (Weinberg \& Katz 2002; Holley-Bockelmann, 
Weinberg \& Katz 2005), although this latest point is a matter of controversy 
(Sellwood 2003; McMillan \& Dehnen 2005).

Baryons have been shown to modify both the radial density profile 
and the triaxiality of the DM halo --- dumping baryons within the halo dilutes
the triaxial potential (Dubinski 1994; Kazantzidis et al. 2004)
and clumpy baryons can level off the density cusp (El-Zant, Shlosman \&
Hoffman 2001). However, the halo triaxiality (at each radius) can also decrease 
during the violent phases of major mergers as they come from random directions 
and during the quiescent phases of slow accretion (outside the halo cusp) --- 
this type of halo evolution is actually implemented in the current work. 

The fraction of stellar bars at cosmological distances is of course an interesting
parameter which can provide us with clues to the evolution of galaxies over the
Hubble time. The original claim that the observed bar fraction decreases sharply 
over redshifts $z \gtorder 0.5$ (Abraham et al. 1999) has been questioned first because
of the specific method of bar detection used in this work (Jogee et al. 2002a,b).
More broadly, analysis of the {\it HST} GEMS\footnote{Galaxy Evolution from Morphology 
and SEDs
(Rix et al. 2004)} survey of about 1,600 galaxies has shown that the fraction  
of strong bars in the optical stays unchanged to $z\sim 1$, i.e., lookback time 
of 8~Gyr (Jogee et al. 2004),
which was confirmed for much smaller samples (Sheth et al. 2003; Elmegreen et al. 2004).
Moreover, Jogee et al. have concluded that the distributions of bar lengths and
their axial ratios is largely preserved as well --- with a corollary for disk
and DM halo properties. Overall, this means that since $z\sim 1$ at least, the galaxy
evolution is probably not driven by major mergers but entered the quiescent phase.

This result can be coupled to a more general issue of the DM halo shape evolution.
El-Zant \& Shlosman (2002) used the method of Liapunov exponents to calculate the
fate of bars in mildly triaxial {\it rigid} halos. They found that the bars cannot
be sustained in these configurations due to the generated chaos.  
Berentzen, Shlosman \& Jogee (2006) have extended this work to include the
self-consistent models with live triaxial halos, i.e., flat and prolate.\footnote{i.e., 
$b>c$ and $a>b$, respectively, where $a, b$ and $c$ are the halo principal axes.} In 
all cases, bars which formed as the initial disk response to the halo
prolateness have dissolved over $\sim$~Gyr timescale, i.e., few bar rotations. 
The series of models with increasing cuspiness in the halo have shown that the
chaos generated by bars tumbling in triaxial halos can also wash out the halo
prolateness and thus the bar may survive. But it remained unclear whether the decrease 
in the
halo triaxiality detected by Berentzen et al. is related to the halo relaxation
because of the insertion of a massive axisymmetric disk or it is the result of
the response in the disk to the halo asymmetry. Our present work provides a compelling
answer to this question.  

The disk formation within a DM halo has been studied before under a long list of 
simplifying assumptions (e.g., Sommer-Larsen et al. 
2003; Immeli et al. 2004; Governato et al. 2006). Here we focus on the disk-halo 
interaction and 
its effects on the developing disk morphology and the feedback onto the halo shape. 
To follow the disk-halo evolution in a more realistic environment, we monitor the
halo assembly in the cosmological framework starting with a $2.5\sigma$ linear 
perturbation in the expanding universe, using the constrained realizations algorithm
(Hoffman \& Ribak 1991) to prescribe the initial conditions. We insert a small `seed'
stellar disk of about 10\% of its final mass within the halo which experiences
a quiescent phase of its dynamical evolution. The disk then grows substantially 
in size and in mass via self-similar mass addition over time periods of $\sim 1-3$~Gyr.
We analyze the subsequent evolution of such a disk-halo system and provide a
comparison with simplified versions of this system. Those include isolated, axisymmetric
and triaxial halos with non-evolving light and massive disks.  

This paper is structured as following: Section~2 deals with the details of numerical
modeling. Results are given in Section~3 and discussed in Section~4.

\section{Numerical Modeling}

We have used the updated version FTM-4.4 of our hybrid code (e.g., Heller \& Shlosman 
1994; Heller 1995) with the total $N\!\sim \!1.3\times 10^6$ particles, of which 1.2M 
represent 
the collisionless DM and 100K --- the collisionless baryonic components, respectively.
The gravitational forces have been computed using Dehnen's (2002) {\tt falcON}
force solver, a tree code with mutual cell-cell interactions and complexity
{\it O(N)}. It conserves momentum exactly and is about 10 times faster than
optimally coded Barnes \& Hut (1986) tree code.
The  gravitational softening of 200~pc to simulate both the evolution of the DM 
component and that of the baryons in the disk has been implemented.
 
We have introduced the following dimensionless model units. The spatial
distance unit is taken as $r\!=\!10$\,kpc, the mass unit is
$M\!=\!10^{11}~{\rm M_\odot}$ and the gravitational constant is chosen
to be ${\rm G}\!=\!1$, which results in a time unit of
$\tau\!=(r^3/G\,M)^{1/2}=4.7\times 10^7$\,yrs. The dynamical timescale, 
$\tau_{\rm dyn}$ within 10~kpc is $\sim 4\times 10^7$~yr for the pure DM halo. In 
these units, the velocity is given in $208~{\rm km~s^{-1}}$. The actual physical 
units are mostly used here.

\subsection{Initial conditions}

We have used the constrained realizations technique (Hoffman \& Ribak 1991; see 
Section~2.1.1 below) to construct a DM halo, and followed its assembly in the 
expanding Friedmann universe. At redshift $z=3$, we have introduced a live `seed' 
stellar disk into the halo and grew it over the next $\sim 1-3$~Gyr, depending on 
the model.

Vacuum boundary conditions and physical coordinates have been used ignoring the cosmological
constant term and assuming the open CDM (OCDM) model with $\Omega_0=0.3$, $h=0.7$ and
$\sigma_8=0.9$. Here $\Omega_0$ is the current cosmological density parameter, $h$
is the Hubble constant normalized by 100~km~${\rm s^{-1}~Mpc^{-1}}$,  and  $\sigma_8$ is 
the variance of the
density field convolved with a top-hat window of radius $8h^{-1}$~Mpc used to normalize
the power spectrum (see also Romano-Diaz et al. 2006a,b). Because we are interested in the
evolution on the sub-galactic scales, the difference with the $\Lambda$CDM cosmology
is small and our results should also stand for the latter. The FTM-4.4 has been
tested in the cosmological context using the Santa Barbara Cluster model (Frenk et al.
1999). Further details are given in Romano-Diaz et al. (2006b).

\subsubsection{The Halo: Constrained Realizations}

The constrained realizations algorithm (Hoffman \& Ribak 1991) was used by applying
linear constraints to specify the initial density field in specified locations and 
evaluated with Gaussian smoothing kernels. The width of a kernel has been fixed
and corresponds to the mass scale on which the constraint has been imposed. Within
the context of the cosmological model and assuming the power spectrum of the
primordial perturbation field, we have generated a constrained realization of the
density field from a random realization of this field. 

The simulations start at redshift $z\!=\!120$, corresponding to $\tau\!=\!0.0127$\,Gyr, 
and ends at 
$z\!=\!0.0$. The DM model has been constructed in such a way as to have a set of 
constraints all located at the origin. The corresponding nested set of perturbations 
has been designed 
to collapse off-center. These density constraints constitute $2.5-3.5\sigma$ perturbations, 
where $\sigma^2$ is the variance of the smoothed field, 
which were imposed on $128^3$ cubic grid with a side of $4h^{-1}$~Mpc. 
The total mass of the DM is $\sim 10^{12}h^{-1}$\,M$_{\odot}$ and it is embedded
in a region corresponding to a mass of $10^{13}h^{-1}$\,M$_{\odot}$ in which the
overdensity is zero of unperturbed Friedmann model.  
The three nested off-center perturbations of $\delta=6.0$, 4.6 and 3.4, corresponding 
to the mass scale of $9.37 \times 10^{10}h^{-1}$\,M$_{\odot}$, $1.87 \times 
10^{11}h^{-1}$\,M$_{\odot}$ and $3.75 \times 10^{11}h^{-1}$\,M$_{\odot}$,  
are designed to collapse by $z_{\rm coll}=5.7,$ 3.8 and 2.2, respectively.  
Over the time of the simulation, the density field has been sampled with 500--1,000 
time outputs.

For the reference, the halo mass within $10$\, kpc is about $1.5$ (or 165K particles),
from $\tau\!=\!38$, or $z\approx3$ (just prior to the seed disk insertion) to
the end of the evolution at $z=0$ ($\tau\!=\!239$). This is about two orders of
magnitude above the halo sampling in cosmological simulations. 

\subsubsection{The seed disk}

The initial `seed' disk is a Miyamoto-Nagai (1975) disk described by the
gravitational potential:
\begin{equation}
  \Phi_{\rm D} = -  \frac{{\rm G}\,M_{\rm D}} {\sqrt{ x^{2}+y^{2}+ \Bl A_{\rm D}
  + \sqrt{B_{\rm D}^{2}+z^{2}} \Br^{2} } } \ ,    
\label{miyamoto}
\end{equation}
where the scalelengths $A_D$ and $B_D$, and the disk mass $M_D$ are given in Table~1.
To include the stellar disk in our models, we first determine the main
principal axes of the triaxial halo, based on its moments of inertia.
We then introduce the disk with its rotation axis being aligned with
the halo minor axis. As we show in Section~3.1, the halo figure is nearly stagnant
over the Hubble time. The disk is placed
with the center of mass position and velocities of the main halo,
iteratively measured in spheres with radii going from $50$\,kpc down to
$0.5$\,kpc. Since the halo is not in equilibrium and
the initial disk mass is negligible compared to that of the halo, we
do not require the halo to relax on the introduced potential. The disk
is inserted at a redshift $z=3$, well after the last major merger (see Section~3).

\subsubsection{Growing a disk within the live halo}

To simulate the process of disk formation in a simplified way, the
disk is gradually grown from the seed, by adding stellar particles to it,
based on the probability given by the Miyamoto-Nagai density distribution.
The central volume density of the disk is kept constant during this time period, 
and we adjust (i.e., increase) the radial and vertical scalelengths of the disk 
correspondingly, keeping the ratio $B_{\rm D}/A_{\rm D}$ constant. Recent
analysis of the GEMS galaxies supports such a disk buildup, when the
surface density does (nearly) not change with redshift while the disk mass
increases with time (Barden et al. 2005). 

Since the equatorial plane of
the disk may change owing to a (slow) inner halo figure tumbling in 3-D, we
determine the instantaneous disk plane from its moments of inertia.  Then, based on
the radial force component at the position of the new particles, we assign
the corresponding circular velocity (without any additional dispersion or
$z$-velocity component (relative to the disk equatorial plane).
 
This method, strictly speaking, does not grow the disk under exact equilibrium 
conditions, but we do not expect this in a realistic disk formation either, 
i.e., most likely it involves some degree of relaxation as well.
We impose a linear mass growth over a specified time period, starting with 
redshift $z=3$ (see Table~2). Overall, the disk mass grows by a factor of
ten over $\sim 1$~Gyr or $\sim 3$~Gyr, which corresponds approximately to
$\Delta\tau=22$ and 60. Because the disk center-of-mass position and the 3-D orientation 
of the disk depend on the 
large-scale motions within the surrounding halo and on the streamers on 
even larger scale in the DM, we have developed a new diagnostics to
address the evolution of disk intrinsic parameters, such as the bar pattern 
speed.

\begin{deluxetable}{lccccr}
\tablecaption{Disk model parameters}
\tablehead{
Model & $M_{\rm D}$ & $A_{\rm D}$ & $B_{\rm D}$ & R$_{\rm cut}$ & $N_{\rm D}$
}
\startdata
{D1}  & 0.0612      & 0.132       & 0.023       & 1.16 &  10\,000  \\
      & 0.6120      & 0.284       & 0.050       & 2.50 & 100\,000  \\[1ex]
{D2}  & 0.1838      & 0.132       & 0.023       & 1.16 &  10\,000  \\
      & 1.8375      & 0.284       & 0.050       & 2.50 & 100\,000  \\
\enddata
\label{table:models}
\tablecomments{Columns: (1) the disk model; (2) mass; (3) radial scalelength;
(4) vertical scalelength; (5) outer disk cutoff; (6) number of particles. The upper 
lines in D1 and D2 correspond to initial and final values of disk parameters, at $\tau_{\rm i}$
and $\tau_{\rm f}$} 
\end{deluxetable}

\begin{deluxetable}{llrrl}
\tablecaption{Simulations}
\tablehead{ 
RUN   & Disk & $\tau_{\rm i}$ [Gyr] & $\tau_{\rm f}$ [Gyr] & Comments
}
\startdata
A\,0  & --   &  --      &  --       &  pure live DM model\\[1.5ex]
A\,1  & D\,1 & 1.81     & 2.84      &  default model\\
A\,1b & D\,1 & 1.81     & 2.84      &  as A1, growing frozen disk\\
A\,2  & D\,1 & 1.81     & 2.84      &  growing thin disk, live halo\\
A\,3  & D\,1 & 1.81     & 2.84      &  growing disk, frozen halo\\
A\,4  & D\,1 & 1.81     & 4.87      &  slowly growing disk, live\\
      &      &          &           &  \hskip .15 in halo\\
A\,5  & D\,1 & --       &  --       &  non-growing disk, live halo\\[1.5ex]
A\,6  & D\,1 & --       &  --       &  low $J$ disk, live halo\\[1.5ex]
B\,1  & D\,2 & 1.81     & 2.84      &  more massive disk, live halo\\
B\,5  & D\,2 & --       &  --       &  non-growing massive disk,\\
      &      &          &           &  \hskip .15 in live halo\\[1.5ex]
C\,1  & D\,1 & 1.81     & 2.84      &  growing disk, live axisym-\\
      &      &          &           &  \hskip .15 in metric halo\\
C\,6  & D\,1 & --       &  --       &  non-growing disk, live axi-\\
      &      &          &           &  \hskip .15 in symmetric halo\\
\enddata
\label{table:runs}
\tablecomments{Columns: (1) the model; (2) the disk model (see Table~1); (3) the start
time of the disk growth; (4) the end time of the disk growth}
\end{deluxetable}

\newpage
\section{Results}

\subsection{Evolution of the pure DM halo --- model A0}

We evolve the halo model from $z\!=\!120$ till $z\!=\!0$, or $\tau\!\approx\!0.013$\,Gyr  and
$11.3$\,Gyr, respectively. We analyze the evolution of a diskless halo in the cosmological 
background for further comparison with halos that harbor light, massive or growing disk 
models. 

\begin{figure}[ht]
\begin{center}
\includegraphics[angle=-90,scale=0.32]{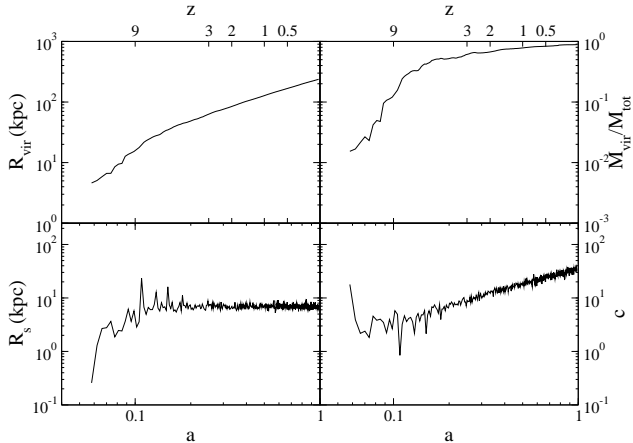}
\end{center}
\caption{Evolution of the NFW scale radius $R_s$, the virial radius $R_{vir}$, the 
concentration parameter $C\equiv R_{vir}/R_s$ and the virial-to-total mass ratio 
$M_{vir}/M_{tot}$, as a function of redshift $z$ and cosmological expansion factor $a$ 
in the pure halo model A0.}
\label{fig1}
\end{figure}
\begin{figure}[ht]
\begin{center}
\includegraphics[angle=-90,scale=0.43]{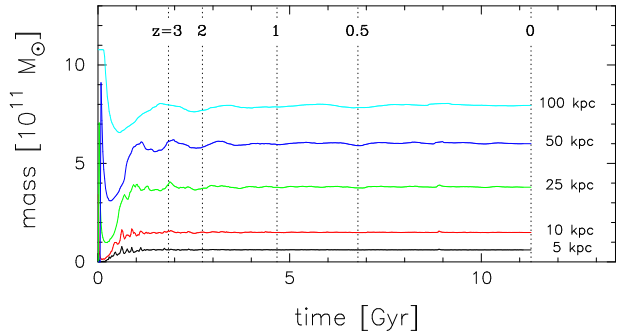}
\end{center}
\caption{Halo mass in the pure halo model A0 within spherical radii and as a function of 
time. The different lines show the mass evolution within spheres of constant radii, 
centered on the main halo. The dashed vertical lines mark, from left to right, redshifts 
$z=3$, 2 and 1, respectively.}
\label{plot3}
\end{figure}
\begin{figure}[ht!!!!!!!!!!!!!]
\begin{center}
\includegraphics[angle=-90,scale=0.43]{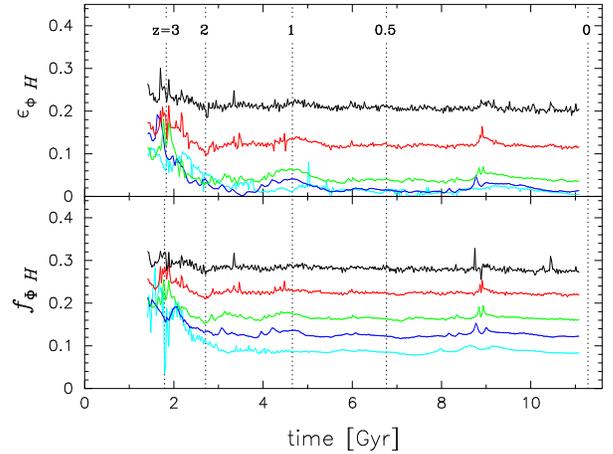}
\end{center}
\caption{Halo shapes based on the {\it isopotential} contours as a function of time in the 
model A0. The different 
colors represent the prolateness, $\epsilon_{\Phi H}\equiv 1-b/a$ (upper panel), and flatness, 
$f_{\Phi H}\equiv 1-c/a$ (lower panel) at different radii (black: 5\,kpc, red: 10\,kpc, green: 
25\,kpc, blue: 50\,kpc, and cyan: 100\,kpc).}
\label{plot2}
\end{figure}
\begin{figure*}[ht!!!!!!!!!!!!!!!!!!!!!!!!!!!!!!!!!!!!!!!!!!!!!]
\begin{center}
\includegraphics[angle=-90,scale=0.73]{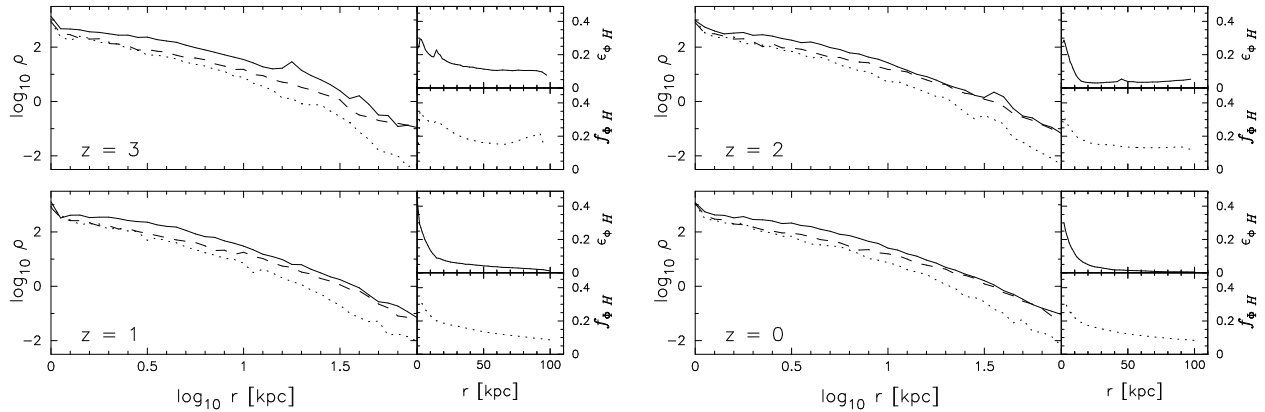}
\end{center}
\caption{Pure DM halo model A\,0. The figure shows the volume density, in units of 
$10^6~{\rm M_\odot~kpc^{-3}}$, along the
 halo principal axes (left panel). The right panels show the halo prolateness
 (upper panel) and its flatness (lower panel). The four frames are taken at different
 redshifts, $z=3$, 2, 1 and 0. The overall halo evolution in the cosmological
 background is shown in the Animation Sequence~1.}
\label{plot4}
\end{figure*}
Initially, the DM expands with the Hubble flow, but detaches and starts its collapse
by $z\sim 15$. The epoch of major mergers ends around $z\sim 7$,
as given by the evolution of the NFW scale radius $R_{\rm s}$ and the halo virial 
radius\footnote{We define the virial radius $R_{\rm vir}$ from 
$M_{\rm vir}=(4\pi/3)\rho R_{\rm vir}^3$, where $M_{\rm vir}$ is the virialized mass 
and the current mean density $\rho=\Delta_{\rm c} \rho_{\rm crit}$, with $\Delta_{\rm c}$
being the current virial overdensity and $\rho_{\rm crit}$ --- the critical density in the 
universe at this time.} $R_{vir}$. These parameters experience abrupt increases during each 
violent event, as shown in Fig.~1 (e.g., Romano-Diaz et al. 2006a,b). 
Following the violent epoch, $R_{\rm s}$ remains constant at $\sim 7$~kpc for the rest 
of the simulations. $R_{vir}$ grows linearly between the mergers and
during the quiescent accretion phase which follows. The latter phase includes the 
minor mergers as well that occur at all redshifts leading to the continuous growth 
of the virialized-to-total halo mass ratio (Fig.~\ref{fig1}). Fig.~\ref{plot3} displays 
the halo mass growth
characterized by the mass assembly within the spherical radii of 5~kpc, 10~kpc, 25~kpc, etc.
The overall halo assembly in the cosmological background is shown in the Animation 
Sequence~1. 
After $z\sim 3$, the central 50~kpc of the halo contains about $650,000$ particles. 
The virialization proceeds from inside out: while the inner 10~kpc 
virialize during the first 1~Gyr, the outer 100~kpc virialize after 2--3 Gyr. The NFW
density profile is established early, by $z\sim 10$. 

The dimensionless cosmological spin parameter for the halo (e.g., Peebles 1969), 
$\lambda\equiv J\, |E|^{1/2}/G\, M^{5/2}$, calculated within $R_{\rm vir}$ 
decreases with time from $\sim 0.1$ at $z\sim 9$ to about 0.013
at $z\sim 1$, or stays a constant, $\lambda_{\rm s}\sim 0.1$ within $R_{\rm s}$,
using notation of Bullock et al. (2001). Here, $E$, $J$  and $M$ are the total energy, 
angular momentum and mass of the halo. These values correspond to the tail of the
average $\lambda$ in numerical simulations (e.g., Barnes \& Efstathiou 1987), because 
we neglect the outer tidal fields which become important at lower redshifts.

For our study of a disk evolution within the live halo, the halo shape, especially its
prolateness, is of a prime importance. Fig.~\ref{plot2} shows the halo prolateness 
$\epsilon_{\Phi H}\!=\!1\!-\!b/a$ and its flatness
$f_{\Phi H}\!=\!1\!-\!c/a$ as a function of time, based on a numerically
stable  ellipse fitting (Halir \& Flusser 1998) of the {\it 
isopotential} contours, for different radii. We note that the isopotentials have a clear 
advantage over the isodensity contours, being much less noisy (Berentzen et al. 2006).
At the same time the values of $\epsilon_\Phi$ and $f_\Phi$ reflect the integral 
properties of the system and have significantly smaller values than those of 
$\epsilon_\rho$ and $f_\rho$ used in the literature and based on fitting the isodensity
contours. 

After about $3$\,Gyr the halo shape evolves in a quiescent manner, largely in response 
to the slow accretion. The halo prolateness and flatness decrease outwards with time because
of our neglect of the large-scale tidal field, i.e., absence of periodic boundary 
conditions. The former becomes small beyond $\sim 25$~kpc radius, while the latter is 
significant at all 
radii. Except for occasional bumps, that reflect the halo 
substructure (i.e., subhalos), both parameters change little with time. It is 
important that the inner halo remains prolate and in a quasi-steady state in this pure halo model.

In Fig.~\ref{plot4} we display four snapshots of the halo volume 
density along its principal axes at different redshifts. These axes have been determined 
using moments of 
inertia. We also show the radial profiles of halo prolateness and flatness. At $z\!=\!3$ 
(or $\tau\!=\!1.81$ Gyr) --- the time of the seed disk insertion, the halo
is prolate, with $\epsilon_{\Phi H}$ just below 0.3 at the center, decreasing to 
below $\sim 0.1$ outside $100$\,kpc. By $z\sim 2$, the halo prolateness is largely washed
out outside the inner 25~kpc, while halo flatness remains significant everywhere.
This detail is important for understanding the evolution of disks immersed
in the halo, specifically for the bar instability --- the {\it outer} halo becomes largely 
axisymmetric in a couple of Gyrs after $z\sim 3$. This happens because of the slow accretion 
of the DM in the absence of the large-scale tidal torques.
However, it is important that the {\it inner} halo, within the central 20~kpc, remains
significantly prolate and hence will interact specifically with the immersed disk.

\subsection{Evolution with growing disks}

\subsubsection{Inserting a disk: the halo relaxation effects}
 
Next, we compare the halo evolution when the disk is immersed in a number of different
ways. The first possibility is when a {\it seed} disk is introduced either abruptly or
more adiabatically at $z=3$. The mass of the seed disk is small, 10\% of 
its final mass and smaller than the halo mass within the disk radius by more than an order 
of magnitude. The disk grows to its final mass within $\sim 1$~Gyr, i.e., $\tau\sim
22$ or in a less intrusive way over a longer period
of time, $\sim 3$~Gyr, i.e., $\tau\sim 60$. To verify that an abrupt positioning 
of the disk within the halo does not produce any visible effects on the halo evolution, 
even more adiabatic insertion of the seed disk is performed
as follows. The disk is brought up gradually over $\Delta\tau = 10$ or 20, keeping the 
disk particle positions frozen, but correcting for the disk center-of-mass with respect 
to the halo. In these test models, the disk reaches its {\it seed} 
mass always at $z=3$ ($\tau\!=\!38$). At this moment we unfreeze the particles and grow the
disk tenfold thereafter. We perform a number of tests where the basic halo parameters are 
compared between the models (see Tables~1 and 2). 

\begin{figure*}[ht]
\begin{center}
\includegraphics[angle=-90,scale=0.65]{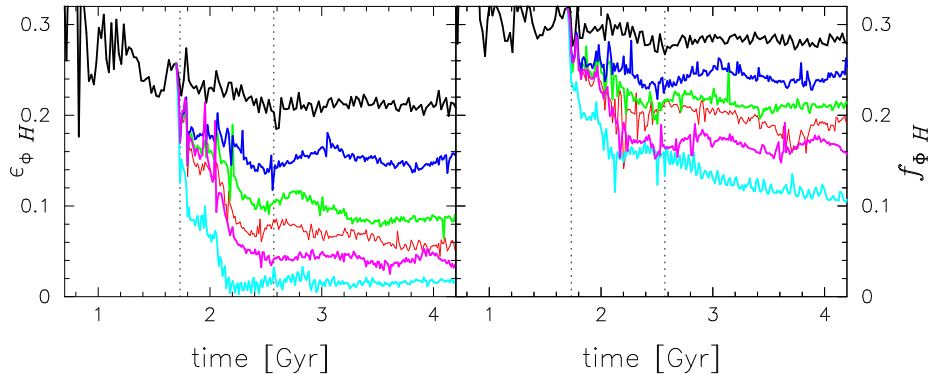}
\end{center}
\caption{Evolution of the halo prolateness (left) and flatness (right) at 5~kpc in different 
model realizations. The model colors are: pure DM halo A0 (black), our benchmark model
with a growing disk A1 (red), slowly growing A4 (green), non-growing A5 (blue), frozen 
growing disk A1b (magenta), massive growing disk B1 (cyan).  Vertical dotted lines are 
$z=3$ and 2 (left-to-right).} 
\label{plot6}
\end{figure*}

We note that the outer part of the halo itself is not in a virial equilibrium, and that the 
seed disk is dominated completely by the halo mass at all radii. The major parameters we 
are concerned with are the halo flatness, $f_{\Phi H}$, and especially its prolateness, 
$\epsilon_{\Phi H}$. We have run a number of test models in order to understand 
to what degree the subsequent evolution of the halo shape depends upon the method of 
disk insertion. We started to bring up the frozen seed disk at $\tau=18$ 
($\sim 0.8$~Gyr) over the
time of 0.47~Gyr and at $\tau=28$ ($\sim 1.3$~Gyr) over the time of 0.94~Gyr. So by $z=3$, 
the halo feels the full seed disk potential, and the disk starts to grow tenfold. In all 
cases the model is evolved for more than 4~Gyr thereafter. We find that various tracks, 
displaying the host halo prolateness and flatness in these test models, are nearly identical 
up to $z\sim 3$, when the disk evolution takes over and the curves separate. 
This means that the diverging evolution we observe in some of the models is not
the result of the halo response to the disk insertion but of other factors, such as
the disk evolution, discussed in the next sections.

Before we discuss in detail the evolution of the key models, we show the halo response to 
the disk for the few most relevant cases: model A5 (a non-growing disk), model A1 
(a standard seed disk growing tenfold), model A4 (same as A1 but a disk growing over 3~Gyr), 
model B1 (a seed disk growing to a massive disk), and model A1b (identical to A1, but the 
disk is growing and remaining in the frozen state). Fig.~\ref{plot6} displays 
$\epsilon_{\Phi H}$ and $f_{\Phi H}$ taken at $5$\,kpc for these models. The most 
dramatic decrease in the halo prolateness we observe in B1, where, unlike in B5, the 
disk grows in mass. This model loses all of its initial prolateness in the inner halo 
over the first $\sim 1$~Gyr. The model A1 loses most of its prolateness as well, while 
A4 mimics its evolution with a somewhat more prolate and flat halo. Interestingly, the model
A1b with the frozen growing disk has a small but measurably larger impact on the halo
shape than the live A1 disk --- to be discussed in Section~4. The model A5 with 
a non-growing seed disk leads to a much smaller decrease in $\epsilon_{\Phi H}$. The pure DM 
halo model A0 is losing less than 20\% of its initial prolateness over the first 1.5~Gyr 
after $z=3$, i.e., almost three times less than the A1. It experiences almost no change 
in the flatness. Note, not only the mass is added in a growing disk but also angular 
momentum, as the new particles possess the maximal allowed $J$ at the radii of their 
insertion. 
 
\begin{figure}[ht!!!!!!!!]
\begin{center}
\includegraphics[angle=-90,scale=0.32]{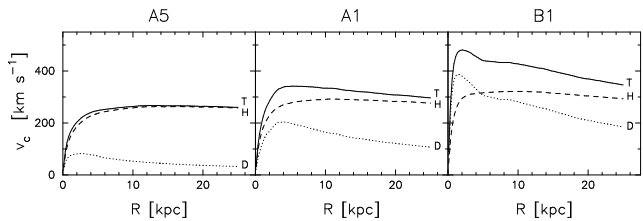}
\end{center}
\caption{Rotation curves for A5 (time $\tau_i=1.81$~Gyr), A1 ($\tau_f=2.84$~Gyr)
and B1 ($\tau_f=4.87$~Gyr) models. Dashed line shows the halo contribution, dotted line
--- the disk contribution, and solid line --- the total rotation curve.}
\label{plot1aa}
\end{figure}
\begin{figure*}[ht!!!!!!!!!!!!!!!!!!!!!!!!!!!!!!!!!!!!]
\begin{center}
\includegraphics[angle=-90,scale=0.6]{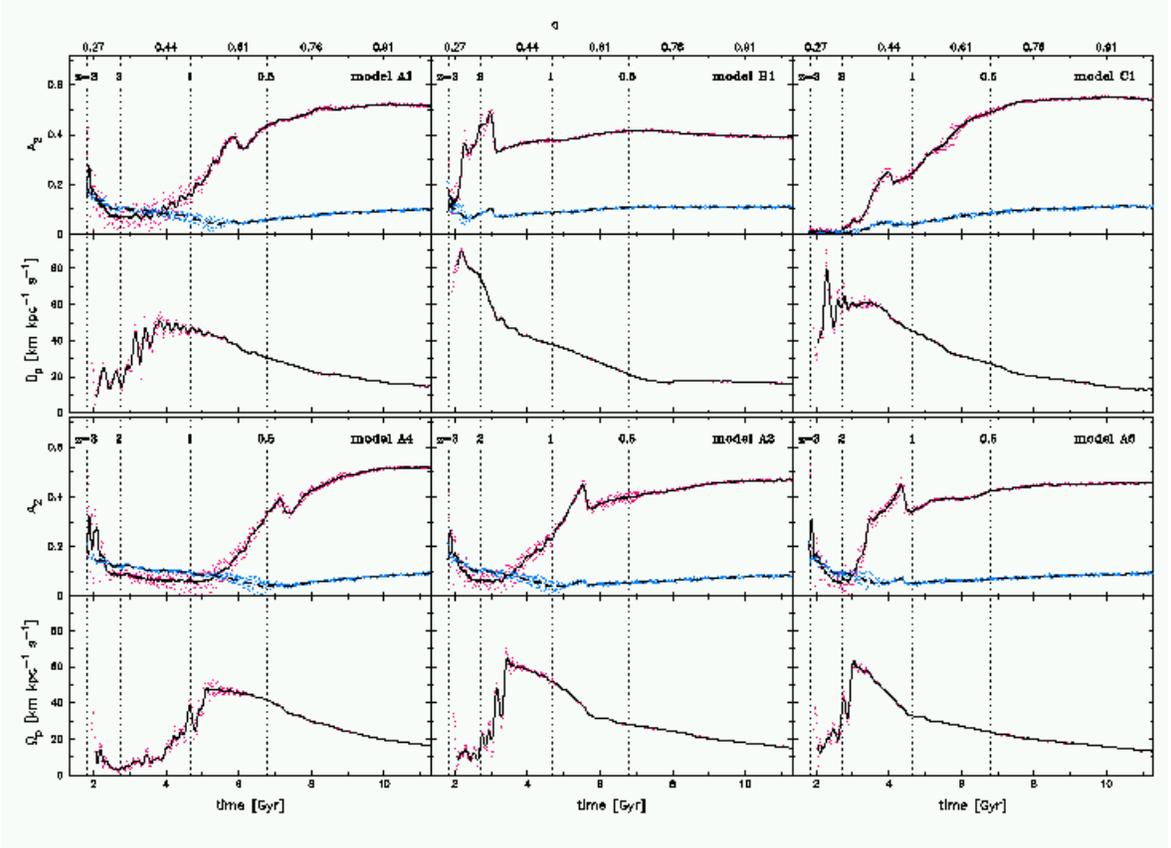}
\end{center}
\caption{Models with growing disks that develop stellar bars. Upper panels: Evolution of 
stellar bars (black line) and DM halo (ghost) bars (dashed cyan line) $m\!=\!2$ Fourier 
amplitudes, $A_2$, integrated within the
central 5~kpc. Lower panels: Bar pattern speeds $\Omega_{\rm b}$. The vertical dotted 
lines mark redshifts $z\!=\!3$ to $z\!=\!1$, respectively. The overall evolution of the 
disk-halo system for the benchmark model A1 is shown in the Animation Sequence~2, where
projections on the individual planes are displayed.}
\label{plot1a}
\end{figure*} 
\vskip .3truein

The rotation curves for models A5, A1 and B1 shown in Fig.~\ref{plot1aa} correspond to the time 
when the disk reaches its maximal mass. The dynamical importance of the disk grows along
this sequence --- while A5 is halo-dominated at all radii, B1 hosts a {\it maximal} disk, 
in the sense of dominating the mass of the inner region. We note that the halo mass is 
dragged inward in 
our models because the disks are inserted nearly adiabatically and initially have 
a negligible mass compared to that of the halo within the same radius. This increase in the 
halo's central mass concentration is an artifact, because of the lack of clumpy baryons in 
our simulations, and leads to an overall increase in the halo cuspiness.

\subsection{Evolution with the growing seed disk --- model A1}

In the {\it initial} response of the seed disk to the triaxial halo, a
bar-like structure forms immediately and fully dissolves
$\sim 1$\,Gyr later. At the dissolution time, the disk has already reached its full
mass. The halo triaxiality gets reduced significantly, compared to
the pure DM model A0, and the halo becomes essentially axisymmetric. Because the disk 
insertion is quasi-adiabatic, the DM cusp is not destroyed but becomes even somewhat 
more pronounced, which is purely an artifact of our way of introducing the disk (see 
Section~3.2.1).

Following the dissolution of the original response to the prolate halo, the
disk stays slightly oval for a prolonged period of time, in excess of $\sim 1$~Gyr. 
For brevity, we define the disk as being barred when the amplitude of $m=2$ mode 
$A_2 \gtorder 0.1$, in all models. Using this convention, the bar in A1 re-appears 
at $\sim 4$~Gyr, this time as a result of a dynamical bar instability. The
details of the bar growth and evolution are shown in  Fig.~\ref{plot1a} and the
Animation Sequence~2 --- the bar experiences
the vertical buckling instability at $\tau \sim 5.9$~Gyr, weakens and continues to
grow secularly, in strength and size, for another $2-2.5$~Gyr thereafter, when 
it saturates.
The pattern speed of the $m=2$ disturbance in the disk is very low initially,
but when the bar re-appears, it
is $\sim 50~{\rm km~kpc^{-1}~s^{-1}}$. This change can be explained by the small 
initial central mass concentration in the model. On the other hand, the fully grown 
disk adds substantially to the central mass and, in addition, drags some of the
DM inward due to the adiabatic way of insertion.

\begin{figure*}[ht]
\begin{center}
\includegraphics[angle=-90,scale=0.73]{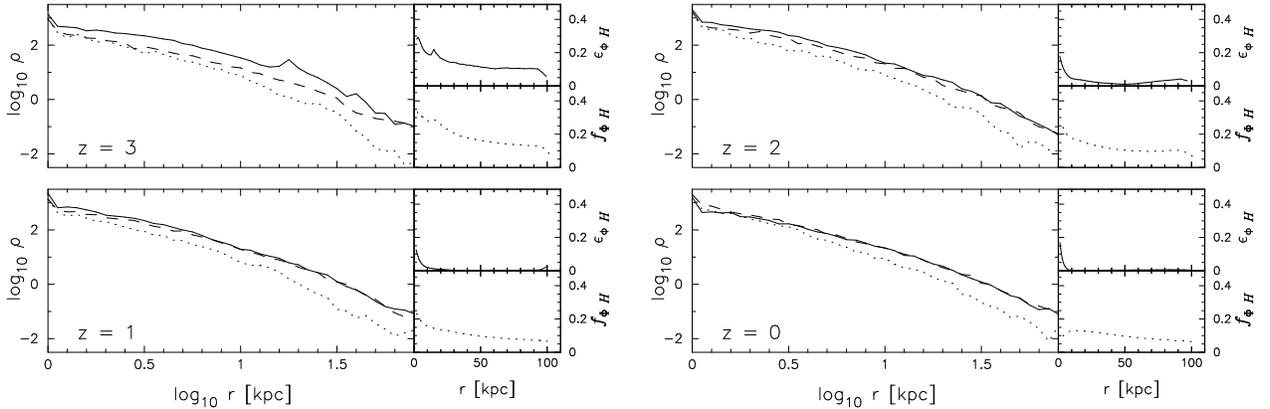}
\end{center}
\caption{The DM halo in the model A\,1.  The figure shows the volume density, in units of 
$10^6~{\rm M_\odot~kpc^{-3}}$, along the
 halo principal axes (left panel). The right panels show the halo prolateness
 (upper panel) and its flatness (lower panel). The four frames are taken at different
 redshifts, $z=3$, 2, 1 and 0.
}
\label{plot4b}
\end{figure*}

The bar in A1 reaches its full strength and size only after the vertical 
buckling, $z\sim 0.5$, but the full disk is in place by $z\sim 2$. What is 
the reason for this delay of $\sim 4$~Gyr, which is comparable to $\sim 10$ disk 
rotations within the bar-unstable region of 10~kpc? Typically, the strong bar appears 
after 2-3 disk rotations, e.g., as in the axisymmetric models LS1--3 of Berentzen et 
al. (2006), if $N$ is sufficiently large and the discreteness noise is 
relatively low. We return to this issue in Section~4.

The initial halo prolateness triggers a response in the disk, with an amplitude, $A_2$, 
comparable to the halo's $m=2$ amplitude, i.e., to the degree of its prolateness. This
response in the disk has a position angle which is normal to that of the halo major
axis and basically represents an oval distortion in the inner disk. Both modes, in 
the disk and the halo, are stagnant, because the halo's spin is negligible, and fade 
away with time. Also, we detect a long-lived spiral structure in the outer disk, in
this and other prolate models, which was also noted by Berentzen et al. (2006). The 
existence of this spirals in a purely collisionless disk within a prolate halo is in 
a sharp contrast with the disk evolution in axisymmetric models. 

As the disk becomes bar unstable, we observe that this bar always excites a `ghost' 
bar in the DM halo (e.g., Athanassoula 2005). This ghost bar is a gravitational wake 
in the halo and not a self-gravitating entity like an actual stellar bar. Its shape differs
from that of the stellar bar -- it is much shorter and `fatter' in all axial ratios.
The ghost bar would dissolve abruptly if the stellar bar would disappear. It 
co-rotates with the stellar bar and has the same pattern speed and nearly the same 
position angle. Aftermath the buckling instability (i.e., $\sim 5.9$~Gyr), the 
stellar bar continues its secular growth, and the amplitude of the ghost bar 
increases as well, although it always remains well below that of the disk bar. 

We have used the Berentzen et al. (2006) method of isodensity fitting to estimate 
the bar size. In A1, it is roughly tracking the bar amplitude by dropping from the
initial 5~kpc (bar semi-major axis) to $\sim 3$~kpc a 1~Gyr later, then increasing
monotonically to about 6.5~kpc, before the secondary buckling 
(Martinez-Valpuesta, Shlosman \& Heller 2006), and leveling off. 


\subsection{Evolution with a growing massive disk --- model B1}
 
We start with a three times more massive seed disk than in A1 (see 
Table~\ref{table:models}). We observed some noticeable differences both in 
the disk and the DM halo evolution as a result. The halo prolateness within
the central 20~kpc drops to near non-existent (see Fig.~5). The initial disk 
response
to the halo is very limited in time --- the onset of the dynamical bar 
instability cuts it off (Fig.~\ref{plot1a}). The bar which appears during 
the disk growth strengthens substantially in less than 0.5~Gyr,
then decays slightly in order to resume its growth for another 1~Gyr.  
It buckles vertically at $\sim 3$~Gyr, which reduces its strength dramatically. 
The bar resumes its secular growth thereafter. For redshifts $z < 0.5$, the 
bar amplitude $A_2$ within the central 5~kpc shows a slight (but monotonic!) 
sign of decrease.  

The $A_2$ amplitude (i.e., prolateness) in the halo decays initially. The
halo ghost bar develops alongside the stellar bar, experiences a sharp drop 
in the amplitude during the stellar bar buckling, and resumes its 
secular growth subsequently. 

The evolution of the pattern speed of the disk bar differs in this model
from that of A1 as well. The stellar bar appears tumbling dramatically
faster at $\Omega_b\sim 90~{\rm km~kpc^{-1}~s^{-1}}$. It decays  strongly
over the evolution time of this model. This behavior characterizes an
efficient $J$-transfer from the disk to the DM halo over the first 6~Gyr
of the bar life and a nearly complete absence of such an interaction in the
next 3~Gyr. 

Furthermore, the bar size follows its $A_2$ variation on the average. The bar 
grows rapidly to above 7~kpc at $\tau\sim 2.3$~Gyr, then drops
to $\sim 5$~kpc over the next Gyr. Interestingly, this drop corresponds
to the sudden change in the bar growth (see above) and is clearly visible in 
Fig.~\ref{plot1a}. For the rest of the evolution, the bar size stabilizes
and decays very slowly in the 8--9~kpc range. All three major characteristics of 
the bar, namely its tumbling rate, bar strength and size, correlate --- all 
remain nearly constant over a prolonged period of time.

\subsection{Evolution with growing disk in axisymmetric halo --- model C1}
  
For a comparison, we have repeated the A1 model evolution in an axisymmetric halo.  
To create this halo, we take a `snapshot' of A0 at $z\!=\!3$ and redistribute 
the particles azimuthally in the random fashion. We kept the original planar radius of
particles unchanged and maintained the halo flattening. Afterwards
we have truncated the halo at $75$\,kpc. We have added the disk to the model and 
relaxed it for $\Delta \tau \!=\! 50$. Finally, we have assigned velocities in the disk 
based on the new axisymmetric halo particle distribution.

\begin{figure*}[ht!!!!!!!!!!!!!!!!!!!!!!!!!!!!!!!!!!!!!!!!!]
\begin{center}
\includegraphics[angle=-90,scale=0.7]{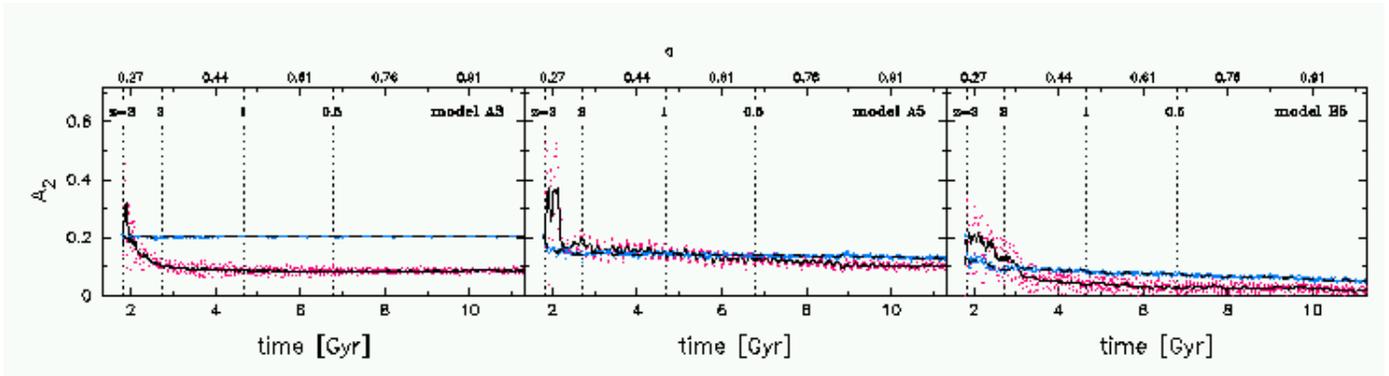}
\end{center}
\caption{Same as the upper frames of Fig.~\ref{plot1a}, but for the models with `failed'
stellar bars.}
\label{plot1b}
\end{figure*}

Naturally, the seed disk-halo system is in dynamical equilibrium at the time of 
insertion, hence no initial disk response to the halo is expected (and observed). 
The bar starts to grow in $\sim 1$~Gyr and buckles subsequently, then resumes its
growth which saturates secularly (Fig.~\ref{plot1a}). The ghost bar behaves similarly
and weakens during the stellar bar buckling, with the follow up growth. 

The stellar bar size grows substantially over the period of $\sim 2$~Gyr. At the buckling
time, the bar size stabilizes around 6~kpc and then experiences a very shallow secular
increase to $\sim 7.5$~kpc, dropping 0.5 kpc at the end of the evolution.


\subsection{Evolution with slowly growing disk --- model A4}
 
In this model we grow the disk over 3~Gyr after its insertion.
The initial response in the disk is the same as in A1 and so is the final structure
of all components (disk, bar and the halo). The halo triaxial shape
is similar as well, with its prolateness remaining slightly larger than in A1 (Fig.~5).
The (initial) halo prolateness is correspondingly more long-lived. 
The main difference between A4 and A1 is that the stellar bar starts to form 
$\sim 2$~Gyr later (Fig.~\ref{plot1a}) --- which is consistent with the similarly
slower disk growth in this model. We relate this delay in the bar growth
to a protracted addition of the angular momentum ($J$) to the disk, as the particles
are added with the maximal $J$ at each energy. To test this, we have run a model
with lower $J$ per added particle, model A6 below.

\subsection{Evolution with a low angular momentum growing disk --- model A6}
 
In order to decrease the angular momentum of the newly added particles to the
growing disk (see model A4), we have re-run the A1 by assigning 
only 85\% of the circular velocity 
for the new particles. The particles have been added only along 
the major axis of the disk (determined from the moment of inertia) in order to
provide some degree of azimuthal coherency to the growing disk. It seems 
that the disk in A6 is more centrally concentrated than in A1 because of the 
large radial excursions of newly added particles with less than maximal $J$.

The resulting disk evolution differs from that of A1 and A4 (Fig.~\ref{plot1a}). 
The stellar bar appears much earlier
at $\sim 3$~Gyr and its nonlinear growth rate is much higher. As a result, the
stellar bar buckles substantially earlier than in A1. However, asymptotically
its evolution converges to that of A1, both in strength and in pattern speed. 
The corresponding initial halo response decays faster as well and the ghost bar
strengthens in tandem with the secular growth of the stellar bar. 

Hence, addition of particles on the lower angular momentum orbits triggers the bar
instability with a higher growth rate, which appears to be sensitive to the 
relatively mild removal of $J$, only 15\%, from the disk. In fact, in A6, the 
stellar bar starts to form immediately after the initial disk response has been 
washed out (Fig.~\ref{plot1a}) --- about 1~Gyr earlier than in model A1. This 
confirms our previous suggestion that our method of adding mass to the 
disk, i.e., with maximal angular momentum at each radius acts as to delay the bar 
instability. To test this, we have run the C6 model, which differs form A6 only
by having an axisymmetric halo. We shall return to this issue in Section~4.

\subsection{Evolution with growing thin disk --- model A2}

Instead of giving the newly added particles {\it vertical} positions based on
the Miyamoto-Nagai distribution function, we have assigned their positions randomly, 
between $z=\!\pm\!100$\,pc. This was done in order to make the growing disk
colder and thus more susceptible to the bar instability. Fig.~\ref{plot1a} shows that the
stellar bar appears somewhat earlier, grows faster, but asymptotically is weaker
than in A1. The halo shape is nearly identical to that of A1.

\subsection{Evolution with growing disk in a frozen triaxial halo --- model A3}

This model was run in order to test the disk response to the undiluted
prolateness of the rigid halo potential. The disk shows a strong
initial response which decays abruptly to some residual amplitude which is
maintained by the halo prolateness. The disk grows as in A1 but no stellar
bar appears over the Hubble time of its evolution (Fig.~\ref{plot1b}). 
There can be a dual reason
for this: either the mass of the disk is so small that the outer disk cannot
absorb enough of the angular momentum of the inner few kpc, or the actual
bar instability is damped by the substantially triaxial halo, as argued
by El-Zant \& Shlosman (2002) and Berentzen et al. (2006).
We have ruled out the former possibility by evolving the same disk in a
frozen axisymmetric halo --- a strong bar has appeared in this case. 
Hence this model confirms both the theoretical predictions and the previous
live numerical simulations that the bar instability is strongly damped by
the halo prolateness, especially in frozen halos which lack the ability to
adjust and to wash out their asymmetry.

\subsection{Evolution with a non-growing disk --- model A5}

We start with the default disk, as in the A1 model, but the seed disk does not grow.
As in A1, a bar-like feature forms due to the initial response to the triaxial halo
and dissolves linearly with time. The residual $A_2$ amplitude both in the
disk and in the DM halo have the same value (Fig.~\ref{plot1b}) and can be simply 
explained as driven by the halo prolateness. Because the mass of the disk remains small
compared to the halo mass within the same radius, the disk has no visible
effect on the halo shape evolution, except that its insertion reduced the halo triaxiality
somewhat, compared to the pure halo model A0. But this effect is substantially smaller
than for the growing disk in A1. The model A5, in fact, tests to what degree
the seed disk affects the halo evolution.

\subsection{Evolution with a more massive non-growing disk --- model B5}

One can suggest, based on the evolution of the B1, that the main reason
for the difference in the bar growth is the initially more massive seed
disk. However this is contradicted by the B5 model, where we use the same
massive seed disk but keep its mass constant. Overall impression of B5 is
that it closely resembles the A5 model (Fig.~\ref{plot1b}). This test model, 
therefore, provides a strong argument that it is not the initial mass of the disk which 
is important but rather its final mass.  
 
\subsection{Bar sizes, disk radial scalelengths and central bulges}

In all of the models, bars appear to evolve only in axisymmetric halos or halos
which lost their prolateness during the disk growth. The timescale of the disk growth
has no effect on the resulting bar size: models A1 and A4 develop nearly identical bars
with semimajor axis of $r_b\sim 8$~kpc and a very similar history. Axisymmetric disk
in C1 leads to the same bar as in A1, and so is the massive disk in B1. The low angular 
momentum disk in A6 results in a somewhat smaller bar of $r_b\sim 5-6$~kpc. This can be 
explained by a more centrally concentrated disk here. Overall, the bar maximal size 
appears to be close to the DM cusp size $R_s\sim 7$~kpc in A0.

The disk growth in our models leads to the increase in the central density. We find,
however, that models that host stellar bars develop much larger central densities
after the period of the bar instability. More massive disks growing adiabatically drag 
in the DM which results in substantially higher central densities, especially in B1. 
Therefore, unlike in Berentzen et al. (2006), we find that the surface density in the 
current disks can be approximated by a single-exponential law plus a central `bulge,' 
instead of a double-exponential law. This `bulge' region corresponds to the 
axisymmetrized stellar bar part of the disk, and we refer here to the bulge only in the 
context of a density fitting.

The surface density in the outer disk, beyond the 
bar, was fit with an exponential profile, $\Sigma(r)=\Sigma_o {\rm exp}(-r/r_d)$, where
$\Sigma_o$ is the normalization constant and $r_d$ is the disk exponential scalelength.
The resulting exponential law has been extrapolated to the center and the excess surface
density was considered as a bulge contribution. The bulge surface density then has been fit 
by the profile $\Sigma(r)=\Sigma_i {\rm exp}[-(r/r_{bg})^{1/n}]$, where $n$---the shape 
parameter corresponds to the pure exponential ($n=1$) or de Vaucouleurs ($n=4$) profiles, 
$r_{bg}$ is bulge's radial scalelength and $\Sigma_i$ its central density (S\'ersic 1968; 
Binney \& Merrifield 1998). For A1, we find that $n\approx 0.8$ during the entire 
simulation, and $r_{bg}\sim 2.3$~kpc when the full disk is in place. 
The scalelength $r_{bg}$ increases somewhat to 2.7~kpc until the bar buckles at 5.9~Gyr 
and then decreases secularly towards 2~kpc, due to the action of the bar which becomes
more centrally concentrated after the buckling instability. The ratio $r_b/r_{bg}$ changes, 
therefore, from $\sim 1.9$ at the buckling to $\sim 4$ at the end of the simulation. 
The exponential scalelength of the outer disk stays around 3.3~kpc after the disk
stops growing.

\subsection{Disk tumbling}

All disk models in our simulations have been permitted to move freely in response
to the streamers in the assembling halo and interactions with halo substructure.
We have detected that the disk plane in model A1 starts to tilt very slowly around an axis 
in its plane about 2~Gyr after the disk insertion (e.g., Animation Sequence~2) and so in 
other models 
as well. This motion has a counterpart in the inner halo, within $\sim 50$~kpc and 
the equatorial planes of the disk and the halo remain aligned over all the length of the 
simulation. The reason for this motion in the halo lies in the streamers excited by the
halo substructure. As a result, we typically do not detect a warping of the disk midplane,
except for some short periods of time. The disk also does not precess for the same reason.

\subsection{Velocity anisotropy in the DM halo}

\begin{figure}[ht!!!!!!!!!!!!!!!!!!!!!!!!!!!!!!!!!!!!!!!!!]
\begin{center}
\includegraphics[angle=-90,scale=0.5]{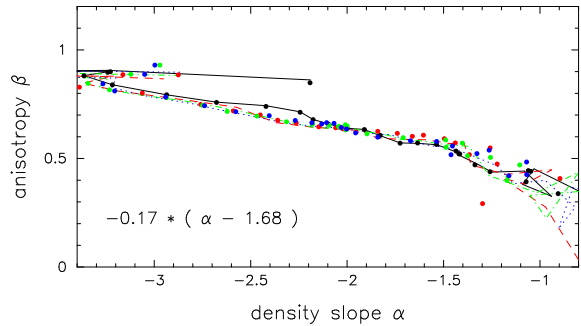}
\end{center}
\caption{Inner dispersion velocity anisotropy parameter $\beta$ vs. density slope parameter 
$\alpha\equiv d{\rm ln} 
\rho/d{\rm ln} r$ for the DM halos in models A0 (lines) and A1 (dots) at different times.
The color coding is as follows: black at 1.81~Gyr, red at 2.84~Gyr, green at 7.09~Gyr and
blue at 11.3~Gyr.}
\label{beta2}
\end{figure}

Hansen \& Stadel (2006) have claimed a correlation between the density slope of the DM halo,
$\alpha\equiv d{\rm ln} \rho/d{\rm ln} r$, and the anisotropy parameter, 
$\beta=1-\sigma_t^2/\sigma_r^2$, where $\sigma_t$ and $\sigma_r$ are tangential and radial
velocity dispersions in the DM. Specifically, they propose a correlation 
$\beta= - 0.2 (\alpha+0.8)$ over about three orders of magnitude in $r$ within the DM cusp, 
with an intrinsic scatter in $\beta$ of 0.05, for slopes with $\alpha > -3$. We have 
measured both $\alpha$ and $\beta$ for two of our models, pure DM model A0 and our 
benchmark model A1. Our best fit for A0 halo is $\beta= - 0.17 (\alpha-1.68)$ for a 
comparable range in $r$ (Fig.~\ref{beta2}). The halo in A1 fits the same relation as in A0.
Hence, we confirm the slope in the correlation proposed by Hansen \& Stadel, but find the
curve being significantly offset from their value. 

\section{Discussion}

We have investigated the evolution of growing stellar disks within assembling asymmetric
DM halos, by means of high-resolution $N$-body simulations. The halos have been constructed 
using constrained realizations in the linear regime and evolved from cosmological initial 
conditions. We have inserted the `seed' disks at the center of 
mass of the innermost halo, at the redshift $z=3$, after the end of the violent 
(merger) phase in the halo evolution. The seed disks, with 10\% of their final mass, 
have been grown over a period of $1-3$~Gyr thereafter. These disks have been allowed to move 
freely within the halo in response to the interactions with the inner and outer 
substructure and with the streamers within the halo. For comparison, we have run 
additional models with more massive disks, non-growing disks, frozen disks, or 
disks in frozen halos, etc.   

Our main result is that a growing disk is responsible for washing out the halo
prolateness and, to a lesser degree, the halo flatness, over a period of time 
comparable to its growth. We find that {\it massive} disks which contribute more to 
the overall rotation curve in the system are more efficient in removing the 
prolateness in the halo. The halo figure rotation remains negligible during this
process. 

In a number of models displayed in Fig.~\ref{plot6}, we show that the halo shape is
very sensitive to the final disk mass, while being reasonably independent of how the
{\it seed} disk is introduced into the system --- abruptly or quasi-adiabatically.
The timescale of the subsequent growth of the disk mass has a small effect on the halo 
shape as well --- such a halo remains slightly more triaxial (model A4). On the other hand, 
when the growing disk was kept frozen, the halo lost slightly more of its prolateness
(model A1b). This has happened because the disk was not able to adjust its shape to that
of the surrounding halo in the latter case. 

The corollary is that the maximal disks most probably reside in nearly axisymmetric 
DM halos. On the other hand, {\it light} disks whose rotation remains halo-dominated at all
radii have progressively smaller effect on the halo shapes and are expected to 
reside in somewhat diluted but prolate halos. Consequently, we envisage that massive 
disks are prone to development of large-scale stellar bars (unless additional adverse 
factors prevail), while light disks have the bar instability damped by the halo triaxiality. 

These corollaries from numerical simulations are complementary and bear interesting 
observational and evolutionary consequences for disk galaxies embedded in DM halos. 
Previous theoretical study (El-Zant \& Shlosman 2002) and numerical simulations 
(Berentzen et al. 2006) have argued that stellar bars and triaxial halos are incompatible,
by-passing the issue of how such systems can form in the first place.
Dubinski (1994) has shown that growing an {\it analytical} spherical 
potential inside a triaxial halos will reduce the halo asymmetry but residual triaxiality
will remain. Kazantzidis et al. (2004) have used cosmological simulations of the DM and
baryonic matter to show that cooling within the halos leads to more spherical halos
than in adiabatic simulations. Models presented here resolve the issue of bar-asymmetric
halo incompatibility by showing explicitly that the growing disk is primarily responsible
for erasing the halo prolateness and diluting its flatness, and that long-lived bars emerge 
only in models which become nearly axisymmetric.
 
Substantial differences in the subsequent evolution of disk models can be detected, 
depending on 
their growth history. This is clearly reflected in the growth rate of the bar instability.
The bar formation is considerably delayed in A1 and proceeds slower in C1 compared to lower 
angular momentum disks in A6 (Fig.~\ref{plot1a}) and C6. The former disks have grown by a 
particle 
addition to the circular orbits, i.e., with the maximal angular momentum $J$ for the 
particle orbital energy. As the bar growth is associated with the loss of $J$ from the 
unstable region, the addition of $J$ should in principle stabilize the disk, temporarily. 
This, however, does not explain why is it that the bar instability is postponed by a 
number of disk rotations after we stop adding mass to the disk in A1 but appears shortly 
thereafter in A6. If this difference arises solely from the additional 15\% of $J$ in A1, 
one should observe the same delay in model C1. However, a substantially smaller delay 
appears in the (axisymmetric halo) C1. 

We note that the original halo prolateness is preserved for a longer period of time in A1 
than in A6, as the direct comparison of {\it halo's} $m=2$ amplitudes in Fig.~\ref{plot1a} 
shows. One observes in this Figure that the stellar bar starts
to grow when the halo's $A_2$ amplitude drops below 0.1, i.e., small $\epsilon_{\phi H}$ 
--- in tandem with calculations 
based on the Liapunov exponents and in self-consistent $N$-body simulations of disk evolution 
in live triaxial halos (El-Zant \& Shlosman 2002; Berentzen et al. 2006). 

We have run a test model (C6) with an axisymmetric halo, as C1, but where particles have
been added with 85\% of $J$ for the circular orbits --- similarly to A6. Its stellar bar 
appears immediately after the disk has reached it final mass, by $z\sim 2$. What is evident 
is that, in both A6 and C6 models, the stellar bar pattern speed has a maximum which is clearly 
larger than in A1 and C1. The simple explanation of this is that models with lower $J$, A6 
and C6, are more centrally concentrated than the latter models and their central dynamical 
timescale is accordingly shorter. Hence, at least in the linear stage, the growth rate of the 
dynamical instability is larger in A6 and C6. It appears that a combination of an addition of 
$J$ and different central concentration govern the bar/disk evolution in the models described 
above. 

All models developing stellar bars also develop {\it ghost} bars in the halo that are 
nearly aligned with the disk bars. Interaction between these bars is a major contributor to
the slowdown of the stellar bar. The ghost bar acts as a gravitational wake and its orbital
structure is fundamentally different from that of the stellar bar --- in particular, the DM 
particles are {\it not} trapped within the ghost bar as they are within the stellar bar. Ghost 
bars
are sensitive enough to respond to the vertical buckling in the stellar bars and they grow
in tandem with the stellar bars, as seen in Fig.~\ref{plot1a}. However, they should not be 
confused with the original halo prolateness which also contributes to the $m=2$ amplitude,$A_2$,
in the halo for the first few Gyrs. We find that this latter contribution serves as a reliable 
indicator for the growth/decay of the stellar bars --- those grow when halo's $A_2$ amplitude 
drops below 0.1.

The density profile in the disk can be fit by a single-exponential profile plus a central bulge
which is somewhat steeper than an exponential. This differs from double-exponential disks that
have developed in Berentzen et al. (2006) models. The reason for this difference appears to be
the quasi-adiabatic insertion of the disk in the current work. This quiescent way of insertion
drags in the DM, without destroying the central cusp. Berentzen et al. (1998) has found a similar
bulge shape for pure collisionless models, which has been modified with the addition of the gas
component in the disk. Overall we find that disk-to-bulge scalelength ratio stays in the range
of $1.2-1.6$, while bar size-to-bulge scalelength ratio increases from about 1.9 to 4. In all 
models, the bar has been roughly confined to within the NFW scalelength $R_s$.

We find that the DM halo cusps are preserved in our numerical simulations, although the quality
of the NFW fit is significantly degraded with the growth of the disks. The reason for this 
apparent robustness of the cusps lies in the quasi-adiabatic insertion of our seed disks which,
in fact, reinforce the cusps by dragging in the DM. Even in models with strong stellar bars, 
the cusps survive. It is possible that in reality they are destroyed by the accreting 
clumpy baryons (e.g., El-Zant et al. 2001, 2004) or by other similar processes (e.g., Tonini, 
Lapi \& Salucci 2006) prior to the disk formation. This issue is beyond the scope of this
work. We also did not address the possible effect that the halo cuspiness has on the halo's
triaxiality (e.g., Berentzen et al. 2006).

How does the evolution of disk-halo systems described in this work fit within the broader
issue of disk galaxy evolution within the cosmological context? While numerical simulations
have shown that halos form as prolate and flattened entities (e.g., Bullock 2002), they appear as nearly axisymmetric (oblate) in the local universe (e.g., Merrifield 2002). On the other hand, 
the high frequency of stellar bars in the local and intermediate redshift universe (Jogee et al. 2004; Elmegreen et al. 2004; Sheth et al. 2003) poses a potential problem because bars and prolate halos are incompatible (El-Zant \& Shlosman 2002; Berentzen et al. 2006). A process which washes
out the halo prolateness, therefore, can be, in principle, responsible for the present
high fraction of barred disks. For this to happen, the reduction of a halo triaxiality should be
achieved {\it quasi-adiabatically}. Otherwise, a galactic disk will be heated up and stabilized
against the bar formation for a prolonged period of time. Major mergers and even frequent minor 
mergers can lead to such an `over-heating' of stellar disks. Furthermore, the major mergers will 
typically destroy the disks.  

A gradual buildup of a galactic disk within an initially {\it a}symmetric DM halo can play the 
role 
of the required quasi-adiabatic process which resolves both issues at once, i.e., the halo 
prolateness and the high observed bar fraction. Jogee et al. (2004) have argued in favor of such a quiescent disk evolution over the last 8~Gyrs, based on the inferred optical fraction of bars and on the distribution of bar ellipticities and sizes over $z\sim 0.2-1$
in recent GEMS survey with the {\it HST}. Moreover, Barden et al. (2005) have shown that the disk mass grows faster with redshift than its surface density. This requires a gradual increase in the disk radial scalelength with $z$ --- a method implemented here to grow the seed disks. 

The scenario that emerges from this {\it gradual} buildup of disks within the triaxial DM halos, in principle, can explain the current population of stellar bars in disk galaxies. Within this
hierarchical merging framework, the DM halos of arbitrary shapes form first during the major
merger epoch. If disks form during this early stage, they are probably destroyed shortly thereafter.
Disks that grow during the next quiescent epoch can be responsible for the washing out of the halo
prolateness and, to a lesser degree, its flatness. The efficiency of this process depends on
the increasing disk-to-halo mass ratio. Large-scale stellar bars will form under these conditions
with a delay --- first bars would form in systems with more massive disks and, therefore, more
oblate halos. For lower redshifts, progressively more disk galaxies will be susceptible to an
intrinsic (spontaneous) bar instability as contrasted by the tidally-induced bars. 

Hence, bars which are observed in the local universe can still be the first generation which formed spontaneously after the host halo prolateness has been washed out. Within this scenario, the intrinsic bar formation is spread out in time. Thus, the observed fraction of stellar 
bars, and distributions of their strength and sizes reflects a steady state between the newly 
forming bars in more axisymmetric halos and bars destroyed by various processes.
The details of the above formation and destruction processes are still sufficiently unclear. For example, we do not know
what is the characteristic timescale of a disk buildup. This work explores the range of $1-3$~Gyrs,
but, in principle, it can be much longer, $\sim 8-10$~Gyrs, as hinted by the results from GEMS, 
and, therefore, can extend beyond $z\sim 2$.

To summarize, we have developed a method to insert seed disks and to grow them subsequently in cosmologically evolving halos in a quasi-adiabatical way. We have investigated the response of the 
halo shape to the growing stellar disk
and the feedback of the halo onto the subsequent evolution of this disk. We find that the
growing disk dilutes the halo triaxiality depending on the final disk mass. Maximal disks,
therefore, are expected to reside in oblate halos, while disks dominated by the halo at all 
radii can reside in prolate halos. Consequently, we expect that the massive disks are prone 
to the bar instability, while light disks have this instability damped by the halo triaxiality.

\acknowledgments
We are grateful to Yehuda Hoffman for producing the constrained realizations, to Clayton 
Heller for converting them into initial conditions for the DM, and to Emilio Romano-Diaz 
for helping with selected diagnostics in Fig.~1. This research has been partially 
supported by NASA/LTSA 5-13063, NASA/ATP NAG5-10823, HST/AR-10284, and by NSF/AST 02-06251 
(to IS).

\clearpage

\end{document}